\documentclass[letterpaper, 10 pt, conference]{ieeeconf}  % Comment this line out if you need a4paper

\IEEEoverridecommandlockouts                              % This command is only needed if
\usepackage[utf8]{inputenc}
\usepackage{dsfont,amsmath,amsfonts,amsbsy,amssymb,amscd}
\usepackage{ntheorem}
\usepackage{bm}  %bold for symbols
\usepackage[version=4]{mhchem}
\usepackage{bigints}
\usepackage{graphicx} % Necessary to use \scalebox
\usepackage{subfigure}
\usepackage{array}
\usepackage{caption}
\usepackage{epstopdf} %This will convert your eps on-the-fly
\usepackage{bigstrut}
\usepackage{comment}
\usepackage{breqn}

\newtheorem{thm1}{Theorem}
\newtheorem{lem1}{Lemma}
\newtheorem{proper}{Property}

\newtheorem{def1}{Definition}
\newtheorem{cond1}{Condition}

\newcommand{\Rset}{\mathbb{R}}
\newcommand{\quotes}[1]{``#1''}
\newcommand\T{\rule{0pt}{2.6ex}}       % Top strut
\newcommand\B{\rule[-1.2ex]{0pt}{0pt}} % Bottom strut
\def\-{\raisebox{.85pt}{-}}
\def\+{\raisebox{.85pt}{+}}
\def\={\raisebox{.85pt}{=}}

\title{\LARGE \bf
A State Observer Design for Simultaneous Estimation of Charge State and Crossover in Self-Discharging Disproportionation Redox Flow Batteries
}

\author{Pedro Ascencio$^*$, Kirk Smith, David Howey, and Charles W. Monroe% <-this % stops a space
\thanks{Pedro Ascencio ({\tt pedro.ascencio@eng.ox.ac.uk}), Kirk Smith ({\tt kirk.smith@eng.ox.ac.uk}), Prof.  David Howey ({\tt david.howey@eng.ox.ac.uk}) and Prof. Charles W. Monroe ({\tt charles.monroe@eng.ox.ac.uk}) are with the Department of Engineering Science, University of Oxford, Oxford OX1 3PJ, United Kingdom.$^*$Corresponding author.}%
}

\begin{document}

\maketitle
\thispagestyle{empty}
\pagestyle{empty}

\begin{abstract}
This paper presents an augmented state observer design for the simultaneous estimation of charge state and crossover flux in disproportionation redox flow batteries, which exhibits exponential estimation error convergence to a bounded residual set. The crossover flux of vanadium through the porous separator is considered as an unknown function of the battery states, model-approximated as the output of a persistently excited linear system. This parametric model and the simple isothermal lumped parameter model of the battery are combined to form an augmented space state representation suitable for the observer design, which is carried out via Lyapunov stability theory including the error-uncertainty involved in the approximation of the crossover flux. The observer gain is calculated by solving a polytopic linear matrix inequality problem via convex optimization. The performance of this design is evaluated with a laboratory flow battery prototype undergoing self-discharge.
\end{abstract}

% adding this as a marker (ignore)
\section{ Introduction }
Many multi-physics models have been developed to improve the design and performance of redox flow batteries (RFBs), in particular for aqueous all-vanadium chemistries \cite{Weber-2011,Xu-2015,Zheng-2014}. More elaborate models include distributed parameter descriptions, for instance considering mass transport by diffusion, convection, and migration, among other dynamics \cite{Shah-2010, Chen-2014,Knehr-2012,Wang-2014}. Alternatively, computationally tractable lumped-parameter models assume pseudo-steady transport of solute across the separator, expressing a proportionality between species fluxes and their concentration differences \cite{Tang-2011,Shah-2011,Kazacos-2012,Boettcher-2016}.

Basic approaches to RFB modelling usually assume perfect reservoir balancing and membrane selectivity, but RFB efficiency and lifetime are both affected by the persistent crossover of active species through the separator \cite{Leung-2012,GMPS-2016,Potash-2016,Xianfeng-2011,Won-2015,Knehr-2012}. Crossover is usually modelled as Fickian diffusion \cite{Schmal-1986,Wiedemann-1998, Sun-2010,Kamcev-2017}, avoiding more complex transport behaviour within the membrane \cite{Luo-2018,Shinkle-2012}. This sort of phenomenological simplification, alongside the assumption that model parameters remain constant during operation\footnote{In general, parameters of these models are assumed known by means of standard experiments or model fits (e.g.\ \cite{Gandomi-2018}).}, can lead to erroneous performance predictions.

To enable more accurate prediction of RFB behaviour and to gain more detailed knowledge about physico-chemical changes, observer-based approaches for lumped parameter models of RFBs have been developed. In some cases such models can provide simultaneous and continuous estimation of the main battery states and parameters \cite{Mohamed-2013,Xiong-2014, Wei-2016, Xiong-2017, Wei-2018}. Generally, these approaches use electrical equivalent-circuit models (ECMs) to emulate the RFB's electrical behaviour, and perform state/parameter estimation with an extended Kalman filter (EKF) \cite{Mohamed-2013,Yu-2014,Xiong-2014,Wei-2018}. ECMs do not explicitly describe the crossover flux that causes self-discharge or irreversible degradation; the consequences of crossover are deduced or detected from the transient observer-predicted variables and parameters.

This paper addresses the simultaneous estimation of battery states and crossover flux for a novel type of RFB based on disproportionation chemistry --- a so-called DRFB \cite{Liu-2009,SM-2019}. The battery is described with a simple isothermal lumped-parameter model, which considers the state of charge in one half-cell (including the associated reservoir) and in one chamber of the reactor, alongside the crossover flux out of the half-cell through the reactor's membrane separator. This paper presents an alternative approach to previous work by the authors \cite{ASMH-2019} which included a general parametric model for the crossover flux with no particular assumptions about a specific transport mechanism through the membrane. The present model instead considers time varying parameters generated by a persistently excited linear system, whose states are included as additional states in the original non-linear DRFB model, and whose space-state representation is suitable for observer design based on the standard Lyapunov second method of stability. The gain of the observer is numerically computed solving the resulting linear matrix inequality (LMI) problem via convex optimization, using a polytopic approach.

\subsection{Notation}
The space of all continuous functions on the domain $\Omega$ into $\mathbb{R}$ is denoted $\mathcal{C}(\Omega; \mathbb{R})$. The set of symmetric positive definite matrices of dimension $n \times n$ is denoted $\mathbb{S}^n_{++}$; $\bm{0}_{n \times m}$ is the null matrix of dimension  $n \times m$ and   $\bm{I}_n$ is the identity matrix of dimension $n \times n$. Also, $\mathbb{R}^+ \equiv [0,+\infty)$, $\mathbb{R}^{m+}=\mathbb{R}^+ \times \ldots \times \mathbb{R}^+$ in $m \in \mathbb{N}$ times,  $\dot{v}(t)\triangleq \frac{dv}{dt}(t)$, $\|x\|\triangleq \sqrt{x^{\top} x}$ with $x \in \mathbb{R}^n$ being a real vector of dimension $n$, and $\|E\| \triangleq \sqrt{\lambda_{\text{max}}(E^{\top} E)}$ with $E$ being a real matrix, where $\lambda_{\text{max}}$ denotes the largest eigenvalue; $\lambda_{\text{min}}$ stands for the smallest eigenvalue.

\section{Modeling DRFBs}
\begin{figure}
    \centering
    \vskip+0.1cm
    \includegraphics[width=\columnwidth]{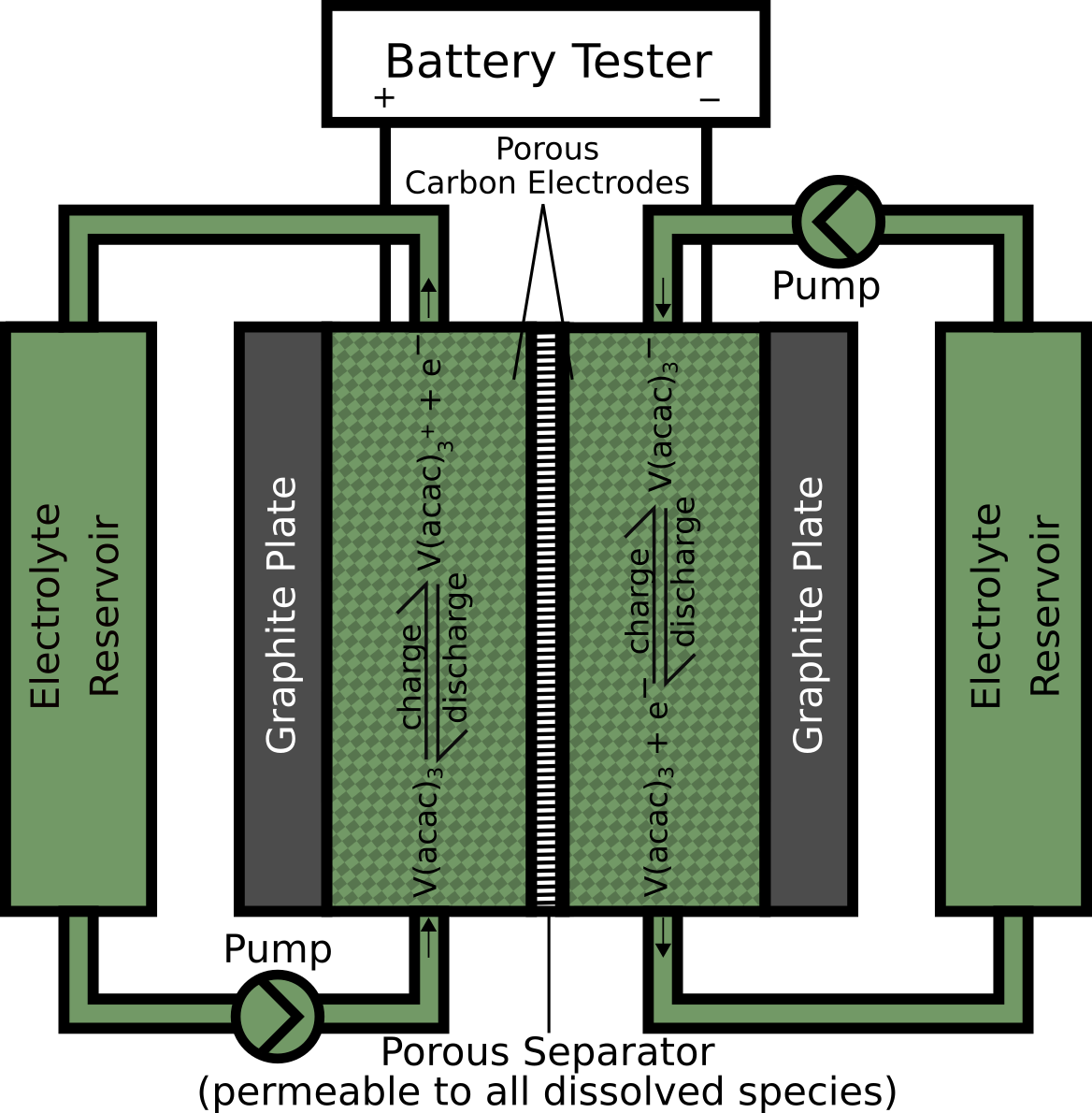}
    \vspace{-0.25cm}
    \caption{\small{Schematic of a DRFB using vanadium acetylacetonate}}
    \label{fig_rfb}
    \vskip-0.25cm
\end{figure}

The DRFB we analyse contains two separated reservoirs storing identical liquid electrolytic solutions in the fully discharged state. The discharged cell contains a single metal electroactive species, vanadium acetylacetonate ($\ce{V(acac)3}$), which can be both oxidized and reduced (see Figure \ref{fig_rfb}). This species has been shown to allow a disproportionation electrochemistry, i.e.\ the battery's charging process causes $\ce{V(acac)3}$ to oxidize on one side of the battery, while it reduces on the other.
A key feature of this DRFB is the use of a nonaqueous solvent, which allows the high redox potential associated with $\ce{V(acac)_3}$ disproportionation. The DRFB configuration may make costly selective separator membranes and periodic active-liquid regeneration unnecessary \cite{SM-2019}, if suitable modes of control can be developed to maintain relatively high energy efficiency during cycling.

In accordance with \cite{Liu-2009}, cyclic voltammetry experiments show two main redox couples associated with $\ce{V(acac)_3}$:

\noindent
\begin{align}
\begin{split}
& \bullet \quad \text{Negative Chamber} \nonumber \\[-0.1cm]
& \ce{  V(III)(acac)_{3} + e- <=>[charge][discharge] [V(II)(acac)_{3}]^{-}}, \\[-0.1cm]
& \bullet \quad \text{Positive Chamber}  \nonumber \\[-0.1cm]
& \ce{  V(III)(acac)_{3} <=>[charge][discharge] [V(IV)(acac)_{3}]^{+} + e^{-} }.
\end{split}
\end{align}

\noindent
Together these establish an equilibrium cell potential of $E^{\text{0}}= 2.18$ [V] \cite{Liu-2009,Shinkle-2012}.  The reversible nature of the reactions makes this type of battery tolerant to crossover electrolyte flux. Crossover causes comproportionation of $\ce{V(II)}$ and $\ce{V(IV)}$ to form $\ce{V(III)(acac)3}$ \cite{Shinkle-2012}, leading to self-discharge, but does not permanently degrade the battery capacity. DRFBs thus enable porous separators to be used, opening up a value tradeoff between the reactor capital cost and its coulombic efficiency \cite{Potash-2016,Liu-2009,Shinkle-2012,James_PhD,SM-2019}.

\subsection{Isothermal Lumped Parameter Model}
\label{model}
In this section, for clarity of explanation, the model formulated by the authors in \cite{ASMH-2019} is described.

Due to the symmetry of the disproportionation chemistry, under identical flows supplied by synchronised pumps, only one side of the battery needs to be analysed to model discharge. In isothermal operation, let  $n$  and  $n_{\text{cell}}$ be the amount of neutral $\ce{V(acac)_{3}}$ species remaining in one reservoir and in the half-cell, respectively. The rate of change of $n$ is due to the current\footnote{Positive current is considered a discharge process.} $I$ driven by the battery, and the crossover flux $Q_x$ through the separator, namely

\noindent
\begin{align}
\label{dn}
& \frac{dn}{dt}(t) = \frac{I(t)}{\mathcal{F}} + Q_x(s(t)),
\end{align}

\noindent
where $\mathcal{F}$ is the Faraday's constant and $Q_x$ is considered to be a function of possible $m \in \mathbb{N}$ variables/states $s$ of the battery, which
could be states, inputs or outputs of the model, or even other measurements, independent of the model: $s: \! \mathbb{R}^+ \! \to \! \Omega \subset \mathbb{R}^{m+}$. Thus, considering the half-cell reactor volume $V_{\text{cell}}$ to be negligible with respect to to the volume in the reservoir $V_{\text{res}}$ and $c(t)\!=\!n(t)/V_{\text{res}}$ being overall concentration of neutral species, the rate of change of the overall system state-of-charge, defined by $SOC(t)\!=\! (n_0-n(t))/n_0 \!=\! \frac{c_0-c(t)}{c_0}$, can be described by

\noindent
\begin{align}
\label{dsoc}
\begin{split}
& \frac{d SOC}{dt} (t) \!=\! \left(\frac{-1}{c_0 V_{\text{res}}}\right) Q_x(s(t)) \!-\! \left(\frac{1}{c_0 V_{\text{res}}\mathcal{F}}\right) I(t),
\end{split}
\end{align}

\noindent
with $c_0\!=\!n_0/V_{\text{res}}$ as initial concentration. Accordingly, based on the principle of conservation of mass applied to one half-cell, assuming perfect mixing (similarly to a continuous stirred tank reactor model) and $c_{\text{cell}}(t)\!=\!n_{\text{cell}}(t)/(\epsilon V_{\text{cell}})$ being the concentration of neutral species in the half-cell, the rate of change of the reactor state-of-charge, defined by $SOC_{\text{cell}}(t)\!=\! \frac{c_0-c_{\text{cell}}(t)}{c_0}$, can be expressed by

\noindent
\begin{align}
\label{dsoc_cell}
& \frac{d SOC_{\text{cell}}}{dt}(t) \!=\! \left(\!\frac{-1}{\epsilon c_0 V_{\text{cell}}}\!\right)\!\frac{dn}{dt}(t) \-\left(\frac{Q(t)}{\epsilon V_{\text{cell}}}\right) \!\Delta SOC(t),
\end{align}

\noindent
with $\frac{d \left(n_{\text{cell}}-n\right)}{dt}(t)\!=\!Q(t) \left( c(t)\!-\! c_{\text{cell}}(t) \right)$, where $Q$ stands for the volumetric flow rate in the reactor (considered measurable and equal in both chambers) and $\Delta SOC(t)\!=\! SOC_{\text{cell}}(t) \!-\!SOC(t)$; $\epsilon$ accouts for the known porosity of the carbon electrode. Therefore, from \eqref{dsoc}-\eqref{dsoc_cell}, the resulting space-state isothermal lumped parameter model for the DRFB is

\noindent
\begin{align}
&   \begin{bmatrix} \frac{d SOC}{dt}(t) \\[0.1cm] \frac{d SOC_{\text{cell}}}{dt}(t)  \end{bmatrix} = \underbrace{\begin{pmatrix} 0 & 0\\ \frac{Q(t)}{\epsilon V_{\text{cell}}} & -\frac{Q(t)}{\epsilon V_{\text{cell}}} \end{pmatrix}}_{A(Q(t))} \begin{bmatrix}  SOC(t) \\ SOC_{\text{cell}} (t)\end{bmatrix} + \nonumber \\
\label{model_x}
& \underbrace{\begin{bmatrix}  -\frac{1}{c_0 V_{\text{res}}} \\[0.1cm] -\frac{1}{\epsilon c_0 V_{\text{cell}}} \end{bmatrix}}_{E} Q_x(s(t)) + \underbrace{\begin{bmatrix}  -\frac{1}{c_0 V_{\text{res}} \mathcal{F}} \\[0.1cm] -\frac{1}{\epsilon c_0 V_{\text{cell}} \mathcal{F} } \end{bmatrix}}_{B} I(t), \\
\label{model_y}
& V_{\text{out}} (t) = \underbrace{E^{\text{0}}_{\text{cell}} \!+\! \frac{2\mathcal{RT}}{\mathcal{F}} \ln{\left(\frac{SOC_{\text{cell}}(t)}{1\!-\!SOC_{\text{cell}}(t)}\right)} \!+\! V_R(t)}_{\Gamma(SOC_{\text{cell}}(t),I(t))},
\end{align}

\noindent
where $V_{\text{out}}$ denotes the output voltage of the battery based on the Nernst equation, $\mathcal{R}$ is the universal gas constant, $\mathcal{T}\!=\!275 \ ^{\circ} \text{K}$ and $V_R=V_R(SOC_{\text{cell}}(t),I(t))$ considers voltage drops due to overpotentials ($V_R(SOC_{\text{cell}}(t),0)=0$).

\subsection{Augmented Parametric Model for Crossover Flux}
The unknown crossover flux in the DRFB can be considered to be a continuous function of the battery variables. Thus, this dynamic can be modelled approximately as the output of a linear system excited by randomly occurring impulses \cite{J-1971}. A particular case of this approach consists in considering the unknown parameter $\theta$ as the output of a $l$-th pure integrator: $\frac{d^l \theta}{dt^l}(t)=0$ \cite{SAF-2000,AS-2001},  namely

\noindent
\begin{align}
\label{Q_app}
\begin{split}
Q_x(s(t)) &= \Psi(s(t)) \theta(t) +  \varepsilon(t), \\
\frac{d \omega}{dt}(t) & = \Lambda(\lambda) \omega(t), \quad \qquad \qquad \qquad \omega(0)=\bm{0}_l,\\
\theta(t) &=[1,0,\ldots,0] \ \omega(t) \\
\Lambda &=\begin{pmatrix} 0 & \lambda_{1} & 0 & \ldots & 0 \\ 0 & 0 & \lambda_{2} & \ldots & \vdots \\
\vdots & \vdots & \vdots & \ddots & \vdots \\ 0 & 0 & 0 & \ldots & \lambda_{l-1} \\ 0 & 0 & \ldots & \ldots & 0 \end{pmatrix},
\end{split}
\end{align}

\noindent
where $\Psi \!\in \!\mathcal{C}(\overline{\Omega} \! \supseteq \! \Omega; \Rset)$ is a continuous and invertible function, $\varepsilon \in \Rset^{m}$ is an unknown dynamic error in the linear parametric model,  and $\theta$ is extended to $l$ states $\omega(t)\!=\![\omega_1(t), \ldots, \omega_{l}(t)]^\top \!:\! \Rset^+ \!\to\! \Rset^{l}$, the dynamic of which is characterised by the matrix $\Lambda \in \Rset^{l \times l}$ for some $\lambda\!=\![\lambda_{1},\ldots,\lambda_i,\ldots,\lambda_{l\-1}]^\top$ with $\lambda_{i} \in \Rset^{+}$ gains of each parameter state.

\subsection{General Approximate Model}
Let $z(t)=[SOC(t), SOC_{\text{cell}}(t)]^{\top}$ be the vector of battery states, $z: \! \mathbb{R}^+ \! \to \! \mathcal{Z} \subset \mathbb{R}^{2+}$.  For the augmented state vector $x(t)=[z^\top(t), \theta(t), \omega_2(t),\ldots,\omega_{l}(t)]^\top \in \Rset^{l\!+\!2}$ ($\theta\!=\!\omega_1$), using the model-approximation of the crossover function \eqref{Q_app}, the DRFB model \eqref{model_x}-\eqref{model_y} can be formulated as

\noindent
\begin{align}
&\dot{x}(t)= A_{e}(\Psi(s(t)),Q(t)) x(t) + B_e I(t)+ E_e \varepsilon(t), \nonumber \\
& y(t)= \Gamma^{-1}(V_{\text{out}}(t),I(t),V_R(t))=\underbrace{[0,\ 1, \ 0, \,\bm{0}_{1 \times l-1}]}_{C_e} x(t),\nonumber \\[-0.3cm]
\label{Emod}
& A_{e}= \begin{pmatrix}  Q(t) A_{\text{cell}} & E \Psi(s(t)) & \bm{0}_{2 \times l-1} \\ \bm{0}_{l-1 \times 2} & \bm{0}_{l-1 \times 1} & M(\lambda) \\ \bm{0}_{1 \times 2}  & 0 & \bm{0}_{1 \times l-1} \end{pmatrix} \in \Rset^{l+2 \times l+2}, \\
& A_{\text{cell}}= \begin{pmatrix} 0 & 0\\ \frac{1}{\epsilon V_{\text{cell}}} & -\frac{1}{\epsilon V_{\text{cell}}} \end{pmatrix}, \nonumber
\end{align}

\noindent
where $x(0)\!=\!x_0$ is the unknown state and parameter initial condition , $y$ is the measurable battery output, $E_{e}\!=\![E; \bm{0}_{l \times 1}] $, $B_{e}\!=\![B; \bm{0}_{l \times 1}] $ and $M\!=\!\text{diag}(\lambda) \in \Rset^{l\!-\!1 \times
l\!-\!1}$ is a diagonal non-singular matrix\footnote{In terms of the observer design, matrices $A_e,B_e,C_e$ and $E_e$ are considered known or part of the design. The inversion of $\Gamma$ is based on Condition \ref{c_inv}, for small errors in the voltage measurements. The symbols  \quotes{,} and \quotes{;} in the  matrix-vector notation denotes a (Matlab type) row and column separator, respectively.}. It is assumed that the augmented model \eqref{Emod} satisfies the following conditions:

\begin{cond1}
\label{c_inv}
The functional structure of the $V_R$ term in \eqref{model_y} is such that the non-linear mapping $\Gamma$ is globally invertible\footnote{In particular, under open-circuit conditions since $V_R(SOC_{\text{cell}}(t),0)\!=\!0$, for $SOC_{\text{cell}}(t) \neq 1, \ \forall \ t \in \mathbb{R}^+$ (monotonic property of the logarithm function).}.
\end{cond1}

\begin{cond1}
\label{c_par}
The battery states $z$ are always bounded: $\sup_t\{\|z(t) \| \} \!\leq\! \gamma_{z}$,  $\forall \ t \in \mathbb{R}^+$, for some $\gamma_{z} \in \mathbb{R}^{+}$. In addition, the crossover flux represents a slow degradation process such that in \eqref{Emod} there exist bounded parameters: $\sup_t\{\|\theta(t) \| \} \!\leq\! \gamma_{\theta}$  and $\sup_t\{\|\omega(t) \| \} \!\leq\! \gamma_{\omega}$  which lead to a bounded approximation error: $\sup_{t}\{\|\varepsilon(s(t)) \| \} \leq\bar{\varepsilon}$ in \eqref{Q_app}, $\forall \ t \in \mathbb{R}^+$, for some $\gamma_{\theta} \in \mathbb{R}^{+}$,  $\gamma_{\omega} \in \mathbb{R}^{+}$ and $\bar{\varepsilon} \in \mathbb{R}^{+}$.
\end{cond1}

\begin{cond1}
\label{c_lip}
 The function $\Psi$ in \eqref{Q_app} is not singular and always bounded: $\sup_t\{\|\Psi(\hat{s}(t))\|\} \leq \! \gamma_{\Psi}$, $\forall \ t \! \in \! \mathbb{R}^{+}$, for $\hat{s}: \! \mathbb{R}^+ \! \to \! \overline{\Omega} \subset \mathbb{R}^{m+}; \overline{\Omega} \supseteq \Omega $,  and for some $\gamma_{\Psi} \in \! \mathbb{R}^{+}$. In addition, this satisfies the Lipchitz condition: $\|\Psi(s(t))\!-\!\Psi(\hat{s}(t))\| \leq \gamma_{\tilde{\Psi}} \|s(t)-\hat{s}(t)\|$, for some $\gamma_{\tilde{\Psi}} \! \in \! \mathbb{R}^{+}$, $ \forall \ s \!:\! \mathbb{R}^+ \to \Omega$. The battery state-variables $s$ are such that $\|\tilde{s}(t)\|=\|s(t)-\hat{s}(t)\| \leq \gamma_{\tilde{s}} \|\tilde{z}(t)\|$  for some $\gamma_{\tilde{s}} \! \in \! \mathbb{R}^{+}$, with $\tilde{z}(t)\!=\!z(t)-\hat{z}(t)$, $\hat{z}: \! \mathbb{R}^+ \! \to \! \overline{\mathcal{Z}} \subset \mathbb{R}^{2^+}, \overline{\mathcal{Z}} \supseteq \mathcal{Z}$, $\forall \ t \in \mathbb{R}^+$.
\end{cond1}

\section{Observer Design}
For the approximate model \eqref{Emod}, the observer design can be carried out using Lyapunov stability theory \cite{Khalil-2002} based on high gain
observer approaches \cite{FHOB-1997,FBH-1998,BFH-1988,NFHF-2002}. The use of the parametric model \eqref{Q_app} and the derivation of the LMI problem aim to improve the tracking performance of the states with a lower observer gain \cite{RH-1994,RSU-1999,AS-2001,ASF-2004}. The objective is to achieve stability in the sense of exponential uniformly ultimately bounded (EUUB) convergence of the observer estimation error  \cite{FHOB-1997}.

\subsection{State Observer}
The proposed high gain-type observer \cite{Gildas-2007} for simultaneous estimation of battery state and parameters for the system \eqref{Emod} has the structure

\noindent
\begin{align}
\label{Eobs}
\begin{split}
\dot{\hat{x}}(t) &= A_{e}(\Psi(\hat{s}(t)),Q(t)) \hat{x}(t) + B_e I(t) + H_{t} \tilde{y}(t) \\
\hat{y}(t) &= C_{e} \hat{x}(t),
\end{split}
\end{align}

\noindent
where $\hat{x}=[\hat{z}^\top, \hat{\theta}, \hat{\omega_2},\ldots \hat{\omega_l}]^{\top}$ and $\hat{y}$ are respectively the estimated augmented state vector (including parameters), and the estimated outputs, $\hat{x}(0)\!=\!\hat{x}_0$ are estimated initial conditions, $\hat{s}$ stands for the estimated states/variables of the battery chosen to model the crossover flux, $\tilde{y}(t)\!=\!y(t)\!-\!\hat{y}(t)$ denotes the output estimation error and $H_t \in \Rset^{l+2}$ is the time-varying observer linear feedback gain.  The following arguments will be instrumental in the observer design procedure:

\begin{def1}
\label{def_00}
By virtue of Condition \ref{c_lip}, the matrix transformation defined by

\noindent
\begin{align}
\label{T}
\hat{T}(t) &= T(\Psi(\hat{s}(t)))= \begin{pmatrix} \bm{I}_{2} & \bm{0}_{2 \times l} \\ \bm{0}_{l \times 2}  & (1/\varrho) \Psi(\hat{s}(t)) \bm{I}_{l} \end{pmatrix},
\end{align}

\noindent
for some $\varrho \in (0,\+ \infty)$, is invertible and it verifies:

\noindent
\begin{align}
\label{T_A}
& \hat{T}(t) A_{e}(\Psi(\hat{s}(t)),Q(t)) \hat{T}^{-1}(t) = \mathcal{A}(Q(t)) , \\
& \qquad \qquad \qquad \qquad \!=\! \begin{pmatrix} Q(t) A_{\textnormal{cell}} & \varrho E & \bm{0}_{2 \times l-1} \\ \bm{0}_{l-1\times 2} & \bm{0}_{l-1 \times 1} & M(\lambda) \\ \bm{0}_{1 \times 2} & 0 & \bm{0}_{1 \times l-1} \end{pmatrix},\nonumber
\end{align}

\noindent
for every $\hat{s} \!:\! \mathbb{R}^+ \to \overline{\Omega}$, $\forall \ t \in \mathbb{R}^+$, so that $ \mathcal{A}$ does not depend on $\Psi$.
\end{def1}

\begin{cond1}
\label{c_T} Based on Condition \ref{c_lip}, the invertible transformation \eqref{T} is always bounded in the following sense:

\noindent
\begin{align*}
& \tau_m\!=\!\inf_{t \in \mathbb{R}^+} \!\left\{\!1\!,\! \frac{\left|\Psi(\hat{s}(t))\right|}{\varrho} \!\right\} \!\leq \!\left\| \hat{T}(t) \right\| \!\leq \!\sup_{t \in \mathbb{R}^+} \! \left\{\!1\!,\! \frac{\left|\Psi(\hat{s}(t))\right|}{\varrho} \!\right\}\!=\! \tau_M,\\
& \left\| \dot{\hat{T}}(t) \hat{T}^{-1}(t) \right\| \leq \gamma_{T} = \sup_{t \in \mathbb{R}^+} \left\{1, \left|\frac{d \Psi(\hat{s}(t))}{d t} \Psi^{-1}(\hat{s}(t)) \right| \right\},
\end{align*}

\noindent
for every $\hat{s} \!:\! \mathbb{R}^+ \to \overline{\Omega}$, for some $\gamma_T>0$, $\tau_m>0$ and $\tau_M>0$,  $\forall \ t \in \mathbb{R}^+$.
\end{cond1}

\begin{cond1}
\label{c_Q}
The volumetric flow rate $Q$ is a bounded measurable variable with $Q_{m}\!=\!\inf_{t} \{Q(t)\}>0$ and $Q_{M}\!=\!\sup_{t} \{Q(t)\}$, so that its domain of operation $\mathcal{Q}\!=\![Q_{\text{m}}, Q_{\text{M}}]$ is a compact set known in advance. Thus, the time-varying matrix $\mathcal{A}$ in \eqref{T_A} can always be embedded in a polytope of matrices:

\noindent
\begin{align*}
\mathcal{A}(Q(t)) \in \mathcal{P}=\mathbf{Co} \{A(Q_{\text{m}}), A(Q_{\text{M}})\}, \\
\end{align*}

\vspace{-0.6cm}

\noindent
$\forall \ t \in \mathbb{R}^+$, where $\mathbf{Co}$ denotes the convex hull (minimal convex polytope) \cite{Boyd-1994,Anstett-2009}.
\end{cond1}

\subsection{Observer Estimation Error}
The dynamical error between the model \eqref{Emod} and proposed observer \eqref{Eobs} can be expressed as the following linear perturbed dynamic model \cite{K-1996}:

\noindent
\begin{align}
\label{Eerr}
\begin{split}
\dot{\tilde{x}}(t) =& \bar{A}_e(t) \tilde{x}(t) + \eta(t),\\
\eta(t) =& \tilde{A}_{e}(t) x(t) + E_{e} \varepsilon(t), \\
\tilde{y}(t) =& C_{e} \tilde{x}(t),
\end{split}
\end{align}

\noindent
where $\tilde{x}(t)\!=\!x(t)\!-\!\hat{x}(t)$ and $\tilde{y}(t)\!=\!y(t)\!-\!\hat{y}(t)$ are estimation errors for states/parameters and outputs, respectively, with $\tilde{x}=0$ being the only stable equilibrium point of the nominal system and $\tilde{x}(0)\!=\!\tilde{x}_0$ are initial error conditions;  $\bar{A}_e(t) \!=\!\hat{A}_e(t)\!-\!H_{t}C_{e}$, $\tilde{A}_{e}(t)\!=\!A_{e}(t)\!-\!\hat{A}_{e}(t)$, with $\hat{A}_{e}(t)\!=\!A_e(\Psi(\hat{s}(t)),Q(t))$ and $A_{e}(t)\!=\!A_e(\Psi(s(t)),Q(t))$. Thus, by meas of the invertible transformation \eqref{T}, given in Definition \ref{def_00}, the observer estimation error \eqref{Eerr}  can be written as:

\noindent
\begin{align}
& e(t) = \hat{T}(t)  \tilde{x}(t), \nonumber\\
& \dot{e}(t) = \left(\hat{T}(t) \bar{A}_{e} \hat{T}^{-1}(t) \right) \! e(t) \! + \! \left(\dot{\hat{T}}(t) \hat{T}^{-1} (t) \right) \! e(t) \! + \! \hat{T}(t) \eta(t), \nonumber \\
& \!\!=\! \underbrace{(\mathcal{A}(Q(t))  \- \hat{T}(t) H_t C_e)}_{\bar{\mathcal{A}}(t)} e (t) \!+\! (\dot{\hat{T}}(t) \hat{T}^{-1}\!(t)) e(t) \!+\! \hat{T}(t) \eta(t), \nonumber \\
\label{Eerr_e}
& \tilde{y}(t) =  C_e \hat{T}^{-1} (t) \tilde{x} (t) =  C_e e(t) ,
\end{align}

\noindent
with  $e(0) = \hat{T}(0) \tilde{x}_0=e_0$, $\eta(t)=\tilde{A}_{e}(t) \hat{T}^{-1}(t) e(t) + E_{e} \varepsilon(t)$ and $\tau_{m}\|\tilde{x}\|^{2}  \leq  \|e\|^2  \leq \tau_{M} \|\tilde{x}\|^{2}$.

\begin{proper}
\label{p_bound}
Given the error dynamic \eqref{Eerr_e}, the expression

\noindent
\begin{align}
\label{delta}
\delta(t) &= \hat{T}(t) \left(\tilde{A}_{e}(t) x(t) (1-\sigma) + E_{e} \varepsilon(t) \right)
\end{align}

\noindent
represents part of the uncertain dynamic which cannot be compensated by the observer feedback gain $H_t$ in \eqref{Eobs} for some $\sigma \in [0,1]$. If Conditions \ref{c_inv}-\ref{c_T} are satisfied, the following upper bound holds:

\noindent
\begin{align}
\label{bardelta}
\| \delta(t) \| & \leq  \sup_{t \in \mathbb{R}^{+}} \{ \| \delta(t) \| \}, \\
& \leq \tau_M \gamma_{E} ( \gamma_{\tilde{\Psi}} \gamma_{\tilde{s}} \gamma_z \max\{\gamma_z,\gamma_{\omega}\} (1-\sigma) + \bar{\varepsilon}) = \bar{\delta}, \nonumber
\end{align}

\noindent
where $\gamma_{E}\!=\! \|E_e\|\!=\!\|E\|$.
\end{proper}

\subsection{Observer Design, Stability and Convergence}

\begin{lem1}
\label{lem_01}
Let $M_1 \in \Rset^{n \times n}$ and $M_2 \in \Rset^{n \times n}$ be arbitrary matrices for some $n \in \mathbb{N}$. If $\alpha  \in \Rset^{+}$ and $\beta \in \Rset^{+}$ are selected so that $\gamma^{2} \leq \alpha \beta$, then the following inequality holds:

\noindent
\begin{align*}
& 2 \gamma \| M_{1} \upsilon(t) \|  \| M_{2} \upsilon(t) \| \leq \upsilon^{T}(t) \left(\alpha M_{1}^\top M_{1} + \beta M_{2}^T M_{2} \right) \upsilon(t),
\qquad
\end{align*}

\noindent
$\forall \ \upsilon: \! \mathbb{R}^+ \! \to \! \mathbb{R}^{n}$  bounded vector signal, $\forall \ t \in \mathbb{R}^+$.
\end{lem1}

\begin{proof}
See details in \cite{RSU-1999,ASF-2004}.
\end{proof}

\begin{lem1}
\label{lem_02}
If  the Conditions \ref{c_par}-\ref{c_lip} are satisfied by \eqref{Emod}, then the following inequality holds:

\noindent
\begin{align*}
\left\|e^\top(t) P \hat{T}(t) \tilde{A}_{e}(t) x(t) \sigma \right\| & \leq \gamma \|P e(t)\| \|\bar{I} e(t)\|,
\end{align*}

\noindent
in terms of its transformed states \eqref{Eerr_e}, where $P \in \mathbb{S}_{++}^{l+2}$, $\bar{I}=\bar{I}_v \bar{I}_v^\top$, $\bar{I}_v=[\bm{I}_{2}; \bm{0}_{l \times 2}]$, $\gamma= \sigma \gamma_E \gamma_{\theta} \gamma_{\tilde{\Psi}}  \gamma_{\tilde{s}}$ $\in \Rset^{+}$, for some  $\sigma$ $ \in [0,1]$.
\end{lem1}

\begin{proof}
Considering that:

\noindent
\begin{align*}
& \tilde{A}_{e}(t) x(t) =  \Big(A_{e}(\Psi(s(t),Q(t))-\Psi(\hat{s}(t),Q(t)) \Big) x(t),\\
&\!\!\!=\![\bm{I}_{2}; \bm{0}_{l \times 2}] [\bm{0}_{2\times 2}, E \tilde{\Psi}(t),\bm{0}_{2 \times l-1}][z\!^\top\!\!(t);\theta(t);\omega_2\!(t);\ldots;\omega_{l}(t)], \\
&\!\!\!=\bar{I} [E \tilde{\Psi} \theta(t);\bm{0}_{l \times 1}],
\end{align*}

\noindent
where $\tilde{\Psi}(t)\!=\!\Psi(s(t),Q(t))\!-\!\Psi(\hat{s}(t),Q(t))$, due to the special form $\bar{I}$, based on the Conditions
\ref{c_par}-\ref{c_lip}, with $\nu(t) =[E \tilde{\Psi}(t) \theta(t);\bm{0}_{l \times1}(t)]$  yield

\noindent
\begin{align*}
e^\top\!(t) P \hat{T}(t) \tilde{A}_{e}(t) x(t) \sigma & =  e^\top\!(t) W \hat{T}(t) \bar{I}  \nu(t) \sigma, \\
\|e^\top\!(t) P \bar{I} \hat{T}(t) \upsilon(t) \sigma \|  & \leq  \sigma \|e^\top\!(t) P \bar{I} \| \|\upsilon(t)\|, \\
& \leq \sigma \|e^\top(t) P \bar{I} \| \|E\| \| \theta(t)\| \| \tilde{\Psi}(t)\|, \\
& \leq \underbrace{\sigma \gamma_E \gamma_{\theta} \gamma_{\tilde{\Psi}}  \gamma_{\tilde{s}}}_{\gamma} \| P e(t)\| \|\bar{I} e(t)\|,
\end{align*}

\noindent
since $\bar{I} \hat{T}(t)\!=\!\bar{I} \!=\! \bar{I} \hat{T}^{-1}\!(t)$, $\|\tilde{s}(t)\| \!\leq \! \gamma_{\tilde{s}} \|\tilde{z}\|=\! \gamma_{\tilde{s}} \| \bar{I} e(t)\|$, for some  $\sigma \in [0,1]$, where $\gamma \in \Rset^+$ accounts for the uncertainty compensated by the observer feedback gain $H_t$ in \eqref{Eobs}, $\forall \ t \in \mathbb{R}^+$.
\end{proof}

\begin{thm1}
\label{teo}
Let Conditions \ref{c_inv}-\ref{c_T}  be satisfied by \eqref{Emod} and $e \in \Omega_r=\{e(t) \in \Rset^{l+2};  \|e(t)\|< r\}$ for some $r \in \Rset^+$. If some positive constants $\alpha$ and $\beta$ are chosen such that $\alpha \beta \geq \gamma^2$ and the solution for the linear matrix inequality problem:

\noindent
\begin{align}
\label{lmi}
& \begin{bmatrix} \-\!\mathcal{A}^{\top}(Q(t)) \! P \- P \mathcal{A}(Q(t))\!+\!C_e^{\top}\!Z^{\top}\!+\! Z C_e \- \beta \bar{I} \-W & \!\!\! \sqrt{\alpha} P \\ \sqrt{\alpha} P & \!\!\! \bm{I}_{l+2} \end{bmatrix}  \!\succeq\! 0,
\end{align}

\noindent
exists for some $P \!\in\! \mathbb{S}^{l+2}_{++}$, $W \!\in\! \mathbb{S}^{l+2}_{++}$, $Z \!\in\! \mathbb{R}^{l+2}$ and $\rho \in (0,1]$, with $\gamma_T \! \leq \! c_W/(2 \rho c_M)$, $\gamma\!=\!\sigma \gamma_E \gamma_{\theta} \gamma_{\tilde{\Psi}}\gamma_{\tilde{s}}$, $c_{m}\!=\!\lambda_{m}(P)$, $c_{M}\!=\!\lambda_{M}(P)$, $c_{W}\!=\!\lambda_{m}(W)$, $\forall \ Q \!\in\! \mathcal{Q}$, then the state observer \eqref{Eobs} using the feedback gain

\noindent
\begin{align}
\label{obs_H}
H_{t} = \hat{T}^{-1}(t) P^{-1} Z,
\end{align}

\noindent
guarantees that the state estimation error \eqref{Eerr} has a EUUB dynamic \cite{FHOB-1997}.
\end{thm1}

\begin{proof}
Let $V \!=\! e^{\top} \! P e$ be a Lyapunov function candidate\footnote{For clarity, the time-dependence in most of the functions after this has been dropped.} with bounds $c_{m} \|e\|^{2}  \leq \! V(e)  \! \leq c_{M} \|e\|^{2}$, where $P \in \mathbb{S}^{l+2}_+$, with $c_{m}\!=\!\lambda_{m}(P)$ and $c_{M}\!=\!\lambda_{M}(P)$.  Its time derivative along the trajectory \eqref{Eerr_e} yields

\noindent
\begin{align}
\label{lyap_var_a1}
& \dot{V} =\dot{e}^\top P e + e^\top P \dot{e},\\
& \!=\!  e^\top \!\left( \bar{\mathcal{A}}^\top(t)  P + P \bar{\mathcal{A}}(t) \right) e +  2 e^\top P (\dot{\hat{T}} \hat{T}^{-1}) e +  2 e^\top P \hat{T} \eta.  \nonumber
\end{align}

If the signal $\eta$ in \eqref{Eerr} is decomposed by a factor $\sigma \in [0,1]$, then $\hat{T} \eta\!=\! \delta_0 + \delta$, where $\delta$ is given in \eqref{delta} and $\delta_0\!=\!\hat{T} \tilde{A}_{e} x \sigma$ represents  the uncertain term which can be compensated by the observer feedback term $H_t y(t)$ in \eqref{Eobs}.  Thus, considering \eqref{obs_H} so that $\bar{\mathcal{A}}(t)\! =\! \mathcal{A}(Q(t))\! -\! P^{-1} ZC_e$, \eqref{lyap_var_a1} can be upper bounded by

\noindent
\begin{align}
\label{lyap_var_b1}
& \dot{V} \leq  e^\top \! \left( \mathcal{A}(Q(t))^\top \! P\!+\! P \mathcal{A}(Q(t)) \! -\! C_e^\top Z^\top \!\!-\! Z C_e \right) \!e +\\
& 2 \underbrace{\left \|e^\top P \hat{T} \tilde{A}_{e} x \sigma \right\|}_{T_1} + 2 \|P e \| \underbrace{ \left\|\dot{\hat{T}} \hat{T}^{-1}\right\|}_{T_2} \|e\| + 2 \|P e \| \underbrace{\|\delta (t)\|}_{T_3}.\nonumber
\end{align}

Regarding term $T_1$, based on Lemma \ref{lem_01} and \ref{lem_02},  the following inequality hols:

\noindent
\begin{align*}
& 2 \left\|e^\top P \hat{T} \tilde{A}_{e} x \sigma \right\|  \leq 2 \gamma \|Pe\|\|\bar{I}e\| \leq  e^\top (\alpha P^\top \! P \!+\! \beta \bar{I}^\top \bar{I} ) e,
\end{align*}

\noindent
for some $\alpha \in \Rset^+$ and $\beta \in \Rset^+$ so that $\gamma^{2} \leq \alpha \beta$. Similarly, with respect to term $T_2$ and $T_3$, considering Condition \ref{c_T} and Property \ref{p_bound}, respectively, an upper bound for \eqref{lyap_var_b3} can be given by:

\noindent
\begin{align}
& \dot{V} \! \leq \!  e^\top \! \underbrace{\left(\mathcal{A}^\top(Q(t)) \! P \+ P \mathcal{A}(Q(t)) \- C_e^\top Z^\top \- Z C_e \+ \alpha P P \!+\! \beta \bar{I} \right)}_{T_4(t)} \! e \nonumber \\[-0.1cm]
\label{lyap_var_b3}
& \qquad \quad  + 2 \gamma_{T} c_{M} \|e\|^{2} + 2 c_{M} \bar{\delta} \|e\|.
\end{align}

Imposing the condition $T_4(t) \!\leq\! -W$, for some $W \in \mathbb{S}^{l+2}_+$ and $\forall \ t \in \mathbb{R}^+$, by means of the Schur complement \cite{Boyd-1994,Cho-1997}, the resulting Riccati-like inequality  \cite{Reif-1999} can be transformed to the constrained time-variant problem \eqref{lmi}, convex in terms of the variables $\{P, Z,W \}$. If this problem is feasible, decomposing $\gamma_{T}$ by a factor $\rho \in (0,1]$, and considering $e \in \Omega_r$, \eqref{lyap_var_b3} can be upper bounded by:

\noindent
\begin{align}
\label{lyap_var_d1}
& \dot{V} \leq  -(c_W\!-\!2 \rho \gamma_{T} c_M) \|e\|^{2} + 2 c_M \underbrace{(\bar{\delta}\!+\! (1\-\rho) \gamma_T r)}_{\Delta} \|e\|, \\[-0.1cm]
& \!\leq \! -(1\!-\mu) \bar{c} \|e\|^2, \forall e \notin \Omega_{\delta}\!=\!\left\{ e \in \Rset^{l+2}; \|e\| \leq r_{\delta}\!=\!\frac{2 c_M \Delta}{ \mu \bar{c}}  \right\}, \nonumber
\end{align}

\noindent
with $\bar{c}\!=\!c_W\!-\!2 \rho \gamma_{T} c_M$, for some $\mu \in (0,1]$, and \eqref{lyap_var_d1}  is negative outside of the region $\Omega_{\delta}$, $\forall \ t \in \Rset^+$. Thus, based on the comparison lemma \cite{K-1996}, the observer estimation error has an exponential convergence  given by: $\|e(t)\| \leq \! \sqrt{\frac{c_M}{c_m}} \|e_0\| \exp\Big(\-\frac{(1-\mu)\bar{c}}{2 c_M}t \Big)$ towards this region (since $P$, $W$ and $\rho$ are such that $\bar{c}\! \geq\!0 \!\Rightarrow \!\gamma_T \! \leq \! c_W/(2 \rho c_M)$) and this remains inside it thereafter, i.e. it is bounded \cite{JZ-1999}. In addition, considering the case of $e \in \Omega_{\delta}$,  the analysis of \eqref{lyap_var_d1} yields

\noindent
\begin{align*}
& \|\tilde{x}\| \leq  \! \sqrt{\kappa_P} \left(\kappa_{\hat{T}} \|\tilde{x}_{0}\| \exp\left(\frac{-\mu \bar{c}}{2 c_M} t \right) \!+\!  \frac{r_{\delta}}{\tau_m} \sqrt{1\-\exp\left(\frac{-\mu \bar{c}}{c_M} t \right)}\right),\\
 & \leq   \max\left\{\sqrt{\kappa_P} \ \kappa_{\hat{T}} \|\tilde{x}_{0}\| ,  r_{\tilde{x}} = \sqrt{\kappa_P} \frac{c_M}{\tau_m} \frac{2}{\mu \bar{c}} \Delta\right\} ,
\end{align*}

\noindent
where $\kappa_P=c_M/c_m$,  $\kappa_{\hat{T}}= \tau_M/\tau_m$, so that the estimation error has an exponential convergence, the ultimate bound value of which is given by $\lim_{t \rightarrow \infty} \|\tilde{x}\|  \leq r_{\tilde{x}}$,  i.e. this has an EUBB dynamic \cite{FHOB-1997}.
\end{proof}

\subsection{Polytopic Design Approach}
Considering that the volumetric flow rate in \eqref{model_x} satisfies Condition \ref{c_Q}, the approximate augmented model \eqref{Emod} admits a polytope description so that the time-variant nature of the LMI problem \eqref{lmi} can be recast as a polytopic convex problem \cite{ASMH-2019,Ascencio-2004,Anstett-2009,Millerioux-2004,Boyd-1994}:

\noindent
\begin{align}
\label{poly_lmi}
& \min_{\bar{\alpha}, \gamma_{Z}}
\left\{ \bar{\alpha} + \kappa_{Z} \gamma_{Z}  \right\} , \\
& \text{subject to:} \nonumber \\[-0.1cm]
& \forall \ i \= 1,2 \! : \! \left\{ \! \! \!
\begin{array}{l}
\begin{bmatrix} \!-\! \mathcal{A}^{\top}_{i} P \!-\! P \mathcal{A}_{i} \!+\! C_e^{\top}Z^{\top} \!+\! Z C_e \!-\! \overline{W} & P \\ P & \bar{\alpha} \bm{I}_{l+2} \end{bmatrix} \succeq 0,
\end{array} \right. \nonumber \\
& \qquad \qquad \ \begin{bmatrix} \gamma_{Z} \bm{I}_{l+2} & Z \\ Z^{\top} & \gamma_{Z} \end{bmatrix} \succeq  0, \nonumber \\
& \qquad \qquad \ \ P=P^{\top} \succ 0, W=W^\top \succ 0, \nonumber\\
& \qquad \qquad \ \ \bar{\alpha} >0, \gamma_{Z} \geq 0, \nonumber
\end{align}

\noindent
with $\mathcal{A}_1 \!=\! \mathcal{A}(Q_m)$ and $\mathcal{A}_2  \!=\! \mathcal{A}(Q_M)$ vertices of the convex hull $\mathcal{P}$; $\bar{\alpha}=1/\alpha$ degree of freedom to maximise the uncertainty allowed by the LMI \eqref{lmi}, for a given $\beta$ and provided its feasible solution $\{P,Z,W\}$, with $\overline{W}=\beta\bar{I}+ W$. The constant $\kappa_{Z}\! \geq \! 0$ is selected to adjust the norm bound of matrix $Z$.

%---------------------------------------------------------------------%
%---------------------------------------------------------------------%
\section{RFB Experimental Methods}
\label{lab}

The experimental setup and methodology has previously been described by the authors in \cite{ASMH-2019}.

Experiments were conducted in an argon-filled glovebox (PureLab, Inert Technologies, USA) using 0.1 M vanadium acetylacetonate (V(acac)\textsubscript{3}, 98\%, Strem, UK) in anhydrous acetonitrile (ACN, 99.8\%, Sigma, UK) dried over molecular sieves (3 \AA, Sigma-Aldrich, USA) with 0.2 M tetraethylammonium tetrafluoroborate (TEABF\textsubscript{4}, Sigma, 99\%, UK) as a supporting salt.

A nonaqueous-compatible flow cell with 2.20 cm\textsuperscript{2} active area was used with reservoir volumes of 18 mL each and a flowrate of 9 mL/min, which corresponds to a reactor residence time of 4.65 s when using electrodes with a porosity $\epsilon$ of 0.87. Impervious bipolar graphite plates (GraphiteStore, USA) were used with porous carbon felt electrodes (Alfa-Aesar, UK) compressed 50\% to a final thickness of 3.17 mm and used in a counter-current flow-through configuration with a porous separator (Celgard 4650, Celgard, USA).

\begin{figure}
    \centering
    \vskip+0.2cm
    \includegraphics[width=6cm,height=6.5cm]{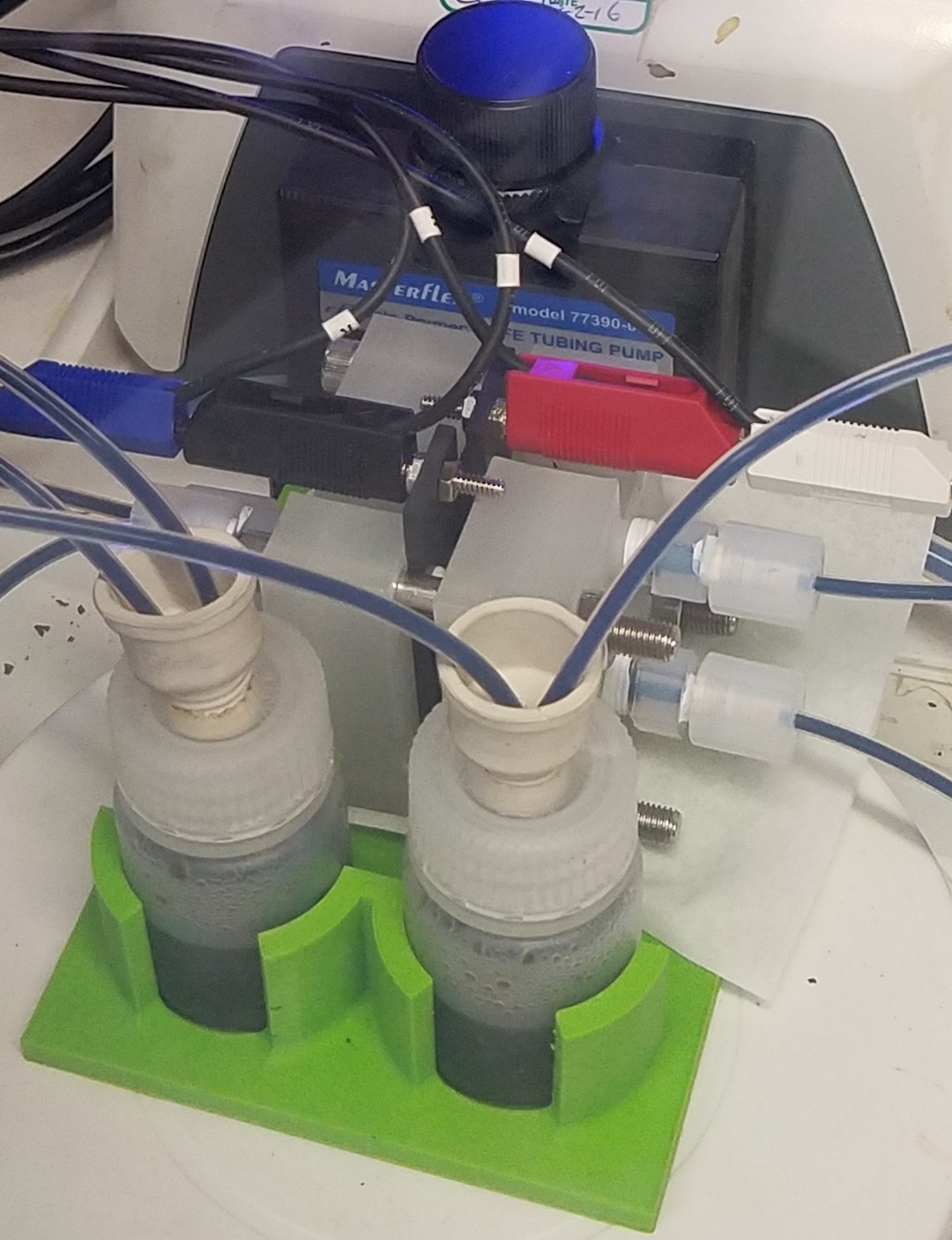}
    %\vskip-0.1cm
     \caption{{\small RFB experimental setup}}
    \label{fig_exp}
    \vskip-0.1cm
\end{figure}

MasterFlex peristaltic pumps (Cole-Palmer, USA) circulated electrolyte through each half-cell with polytetrafluoroethylene (PTFE) tubing using perfluoroalkoxy alkane (PFA) compression fittings. Wetted materials in the system consisted entirely of PTFE, PFA, polypropylene (PP), impervious graphite, and carbon felt. A photograph of the experimental setup is shown in Figure \ref{fig_exp}.

%---------------------------------------------------------------------%
%---------------------------------------------------------------------%
\section{Experimental Results}
\label{results}

As preconditioning procedure, the flow battery used in the experiment was cycled three times at 20 mA/cm\textsuperscript{2} between 3 V and 1 V.  After this, the battery was charged up to the 3 V (voltage cutoff) to let it self-discharge at open circuit. Values of the experimental parameters in the model \eqref{model_x}-\eqref{model_y} are indicated in Table \ref{tconst}.

\begin{figure}[thpb]
\vskip-0.5cm
\hskip-0.25cm
      \begin{tabular}{c}
       \includegraphics[width=9cm,height=6cm]{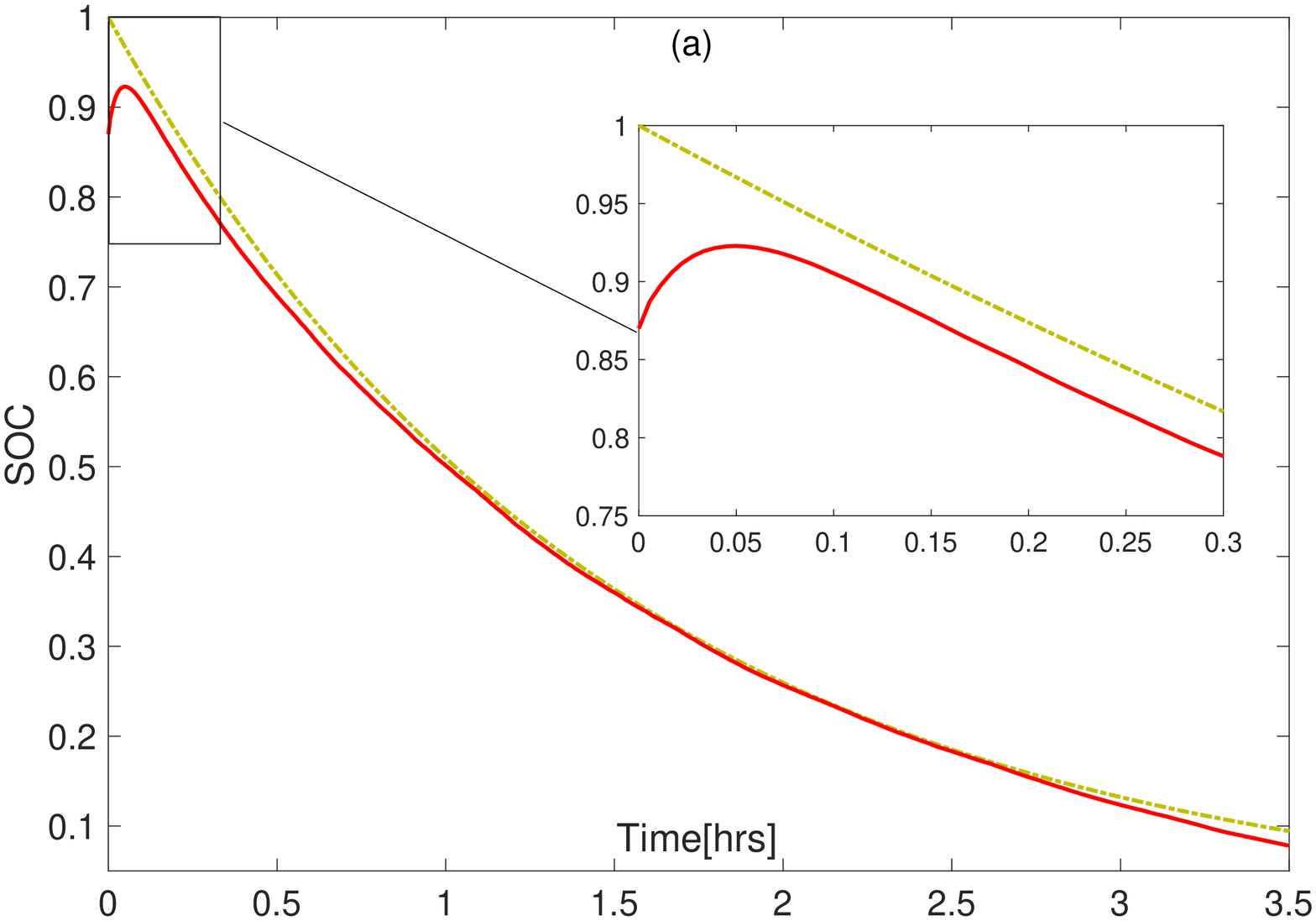} \\[-0.45cm]
       \includegraphics[width=9cm,height=6cm]{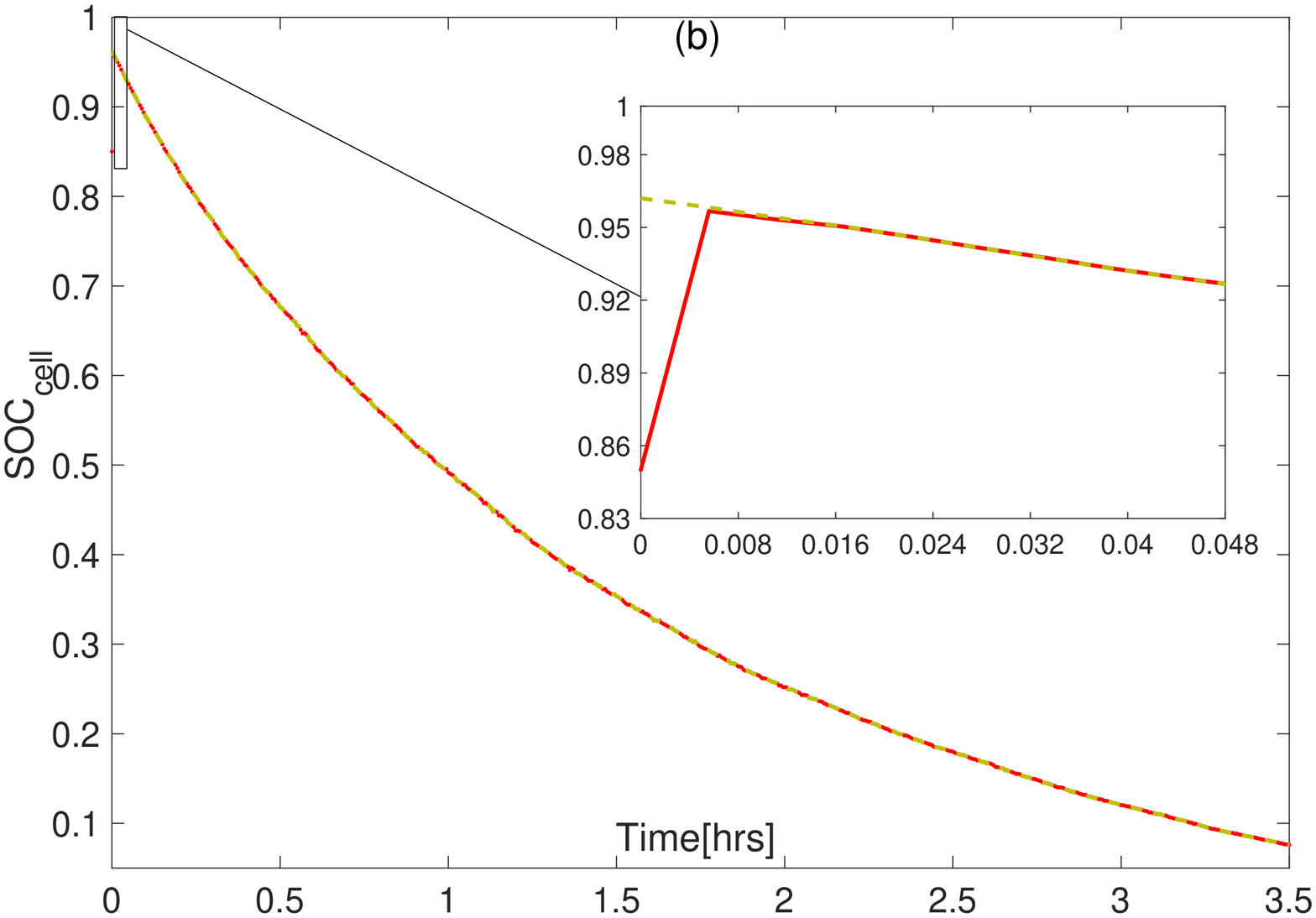}\\[-0.45cm]
       \includegraphics[width=9cm,height=6cm]{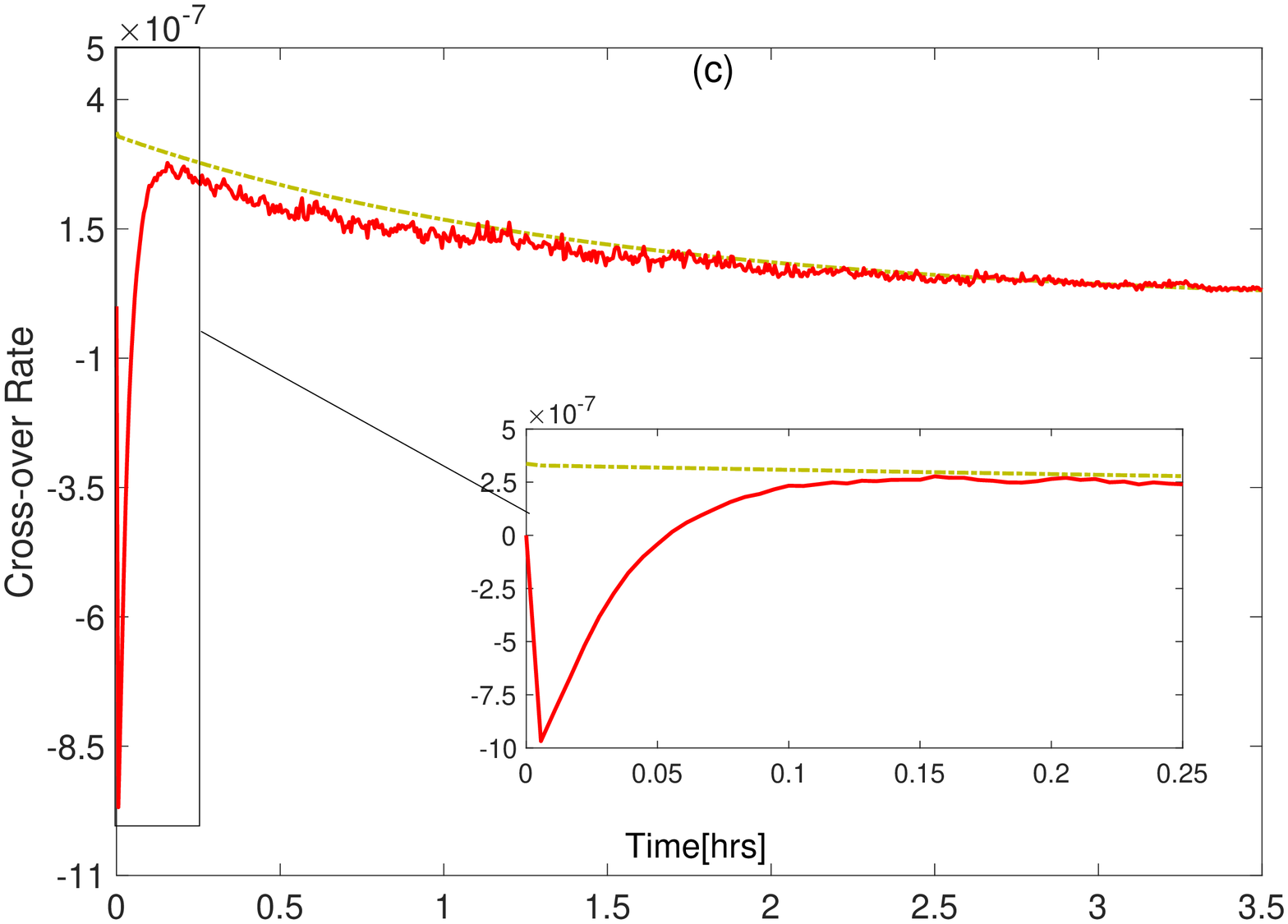}\\[-0.45cm]
       \includegraphics[width=9cm,height=6cm]{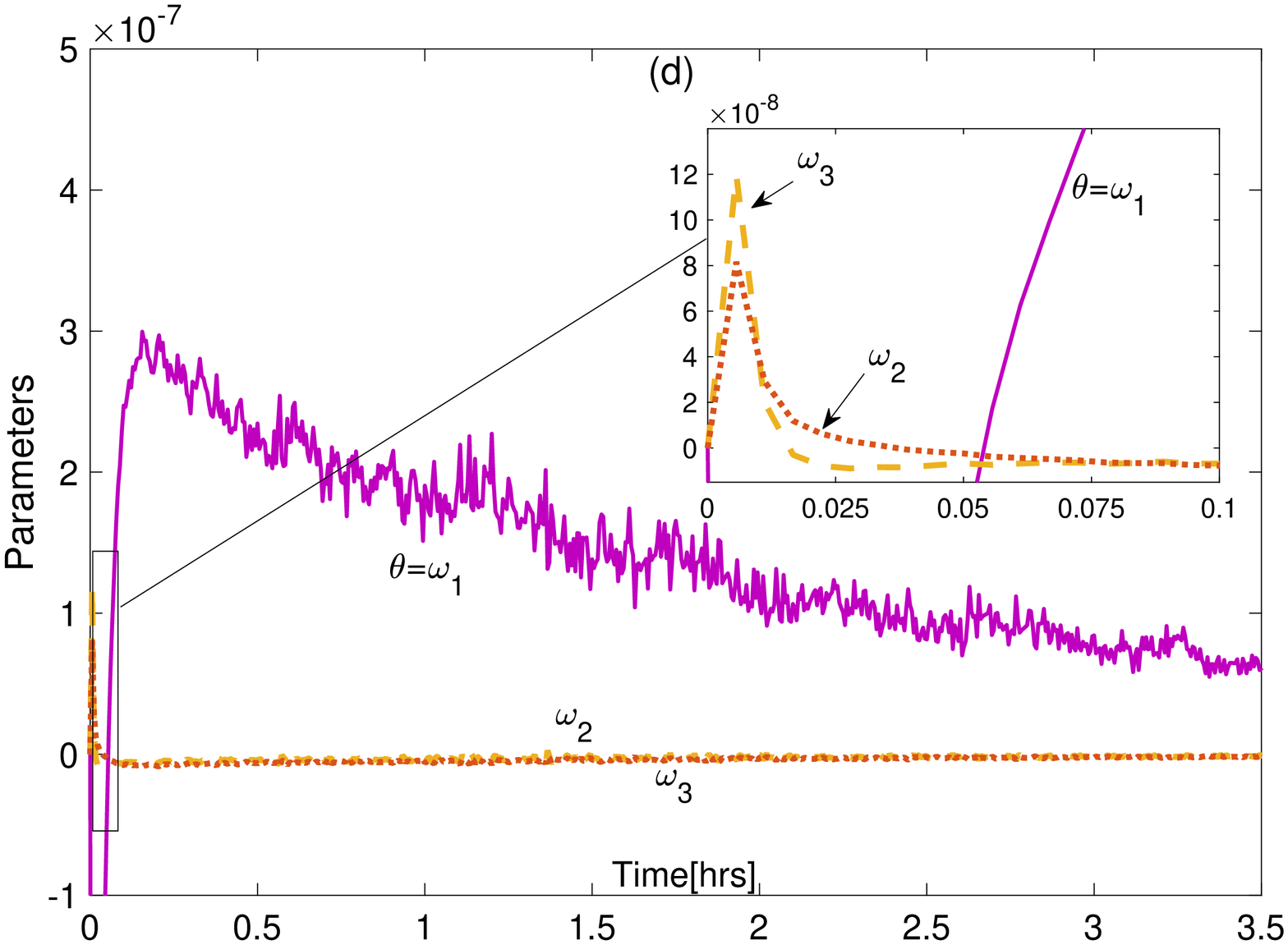}
       \end{tabular}
       \vspace{-0.25cm}
       \caption{\small{Battery model \eqref{model_x}-\eqref{model_y} results with constant parameters in  \eqref{lin_crossover} (dash-dotted lines), alongside augmented state observer estimate (solid lines) for: (a) overall state-of-charge; (b) state-of-charge of the half-cell; (c) crossover flux; (d) estimated parameters for \eqref{Q_app}.}}
       \label{fig_resul01}
\vskip-0.5cm
\end{figure}

\begin{figure}[thpb]
\vskip-0.25cm
\hskip-0.25cm
      \begin{tabular}{c}
       \includegraphics[width=9cm,height=6cm]{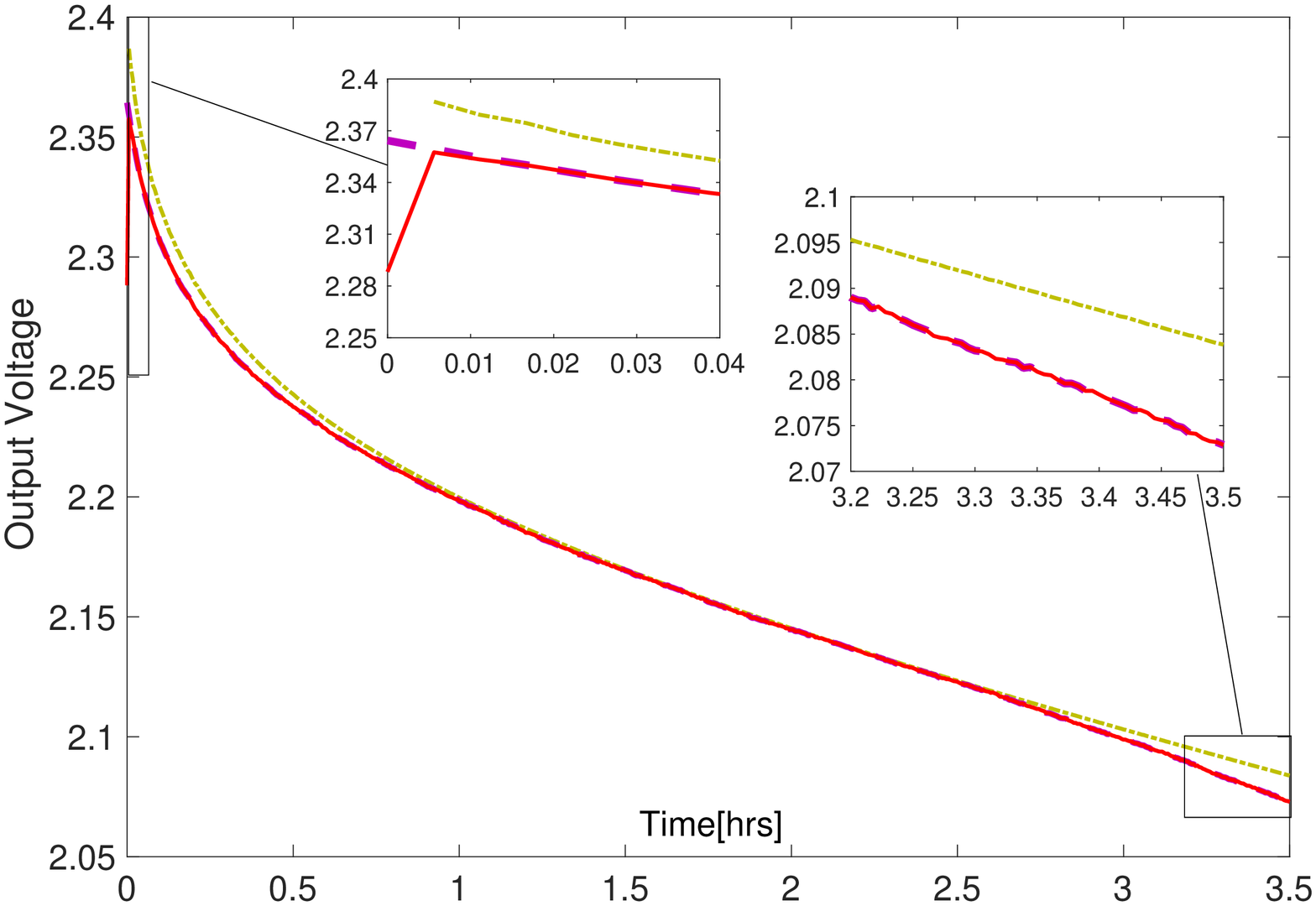}
       \end{tabular}
       \vspace{-0.3cm}
       \caption{\small {Voltages with respect to self-discharge time, including: measured voltage response (dotted line); battery model \eqref{model_x}-\eqref{model_y} with constant parameters in \eqref{lin_crossover} (dash-dotted line); augmented state observer estimate (solid line).
       }}
       \label{fig_resul02}
\vskip-0.15cm
   \end{figure}

For comparison purposes, the following linear function for the crossover flux in \eqref{model_x} has been considered:

\vspace{-0.2cm}

\noindent
\begin{align}
\label{lin_crossover}
Q_x(SOC_{\text{cell}}(t))=k_{\text{mt}} \ c_0 \overbrace{\left(\frac{c_0-c_{\text{cell}}(t)}{c_0}\right)}^{SOC_{\text{cell}}(t)},
\end{align}

\noindent
where $k_{\text{mt}}\!=\!5.6142 \! \cdot \! 10^{\!-8\!}$ L/min is the mass-transfer coefficient. This value for $k_{mt}$ was calculated by fitting the model, \eqref{model_x}-\eqref{model_y}, including \eqref{lin_crossover}, to the experimental data assuming an initial condition of 100\% for $SOC_{\textrm{cell}}$ \cite{ASMH-2019}. Figure \ref{fig_resul01}: (a)-(b) and (c) depicts the battery states and crossover term, respectively (dash-dot lines) and their estimated values by the proposed observer (solid lines), alongside the estimated parameters of the model. Figure \ref{fig_resul02} shows the voltage output predicted by the model during self-discharge (dash-dot line), which is broadly in agreement with its measured value (dot line), excepting zones where $SOC_{\textrm{cell}}$ reaches its extreme values, caused by possibly unmodeled dynamics in the argument of the Nernst equation \eqref{model_y}.

With respect to the proposed on-line augmented state observer \eqref{Eobs}, $\hat{z}_0\!=\![0.87,0.85]^{\top}$ and $\hat{\omega}_0\!=\!\bm{0}_{l \times 1}$ have been selected as its initial conditions. For the crossover approximation \eqref{Q_app}, $s\!=\!SOC_{\text{cell}}\!=\!z_2$, $\Omega \!=\![0,1]$, $\Psi(\hat{z}_2(t))\!=\!0.5(1\!+\!\hat{z}_2(t))$ and a third order pure integrator ($l\!=\!3)$ for the dynamic of the parameters, with $\lambda=[0.5, 0.025]^\top$, have been considered. Regarding the observer gain in \eqref{obs_H}, $\varrho\!=\!10^{\!-\!4}$ has been selected to obtain a value to make the observer less sensitive to noise. The numerical solution of the polytopic LMI problem \eqref{poly_lmi}, considering $Q_{\text{m}}\!=\!0.25Q$, $Q_{\text{M}}\!=\!2 Q$, $\beta=10^{\!-\!4}$ and $\kappa_Z\!=\!0.01$, has been obtained via the Yalmip toolbox for Matlab \cite{Yalmip} using the SDP package part of the Mosek solver \cite{Mosek}. More details about its Matlab code implementation and data can be found in \cite{Matlab_code}.

As it has been corroborated by the authors in \cite{ASMH-2019}, the states estimated by the proposed observer (see Figure \ref{fig_resul01}, solid lines) describe similar behaviour to the model \eqref{model_x}-\eqref{model_y}-\eqref{lin_crossover}. In particular, the estimated crossover flux is in agreement with the linear relationship proposed in \eqref{lin_crossover}, the dynamical parametric model of which \eqref{Q_app} helps to achieve a good tracking performance in the transient response under low magnitudes for the observer gain.

%-----------------------
\begin{table}[t]
\centering
\vskip+0.25cm
\begin{tabular}{ c | m{3.7cm} | l | l}
\hline
{\bf Symbol} & {\bf Description} & {\bf Value} & {\bf Units} \T\B \\
\hline \hline
$V_{\text{res}}$  			& Reservoir volume	 	                            & 17.6 & mL \T \\%$17.6 \cdot 10^{\!-\!3}$     & L \T \\
$V_{\text{cell}}$  		    & Half-cell volume 		                            & 0.6985 & mL \T \\%$9.3654 \cdot 10^{\!-\!4}$   & L \T \\
$c_0$  			            & Initial concentration  $\ce{V(acac)3}$            & $0.1$		                & mol/L \T \\
$Q$  			            & Volumetric flow rate 				                & 9.0 & mL/min \T \\%$1.5 \cdot 10^{\!-\!4}$      & L/s \T\\
%$k_{\text{mt}}$  		    & Mass-transfer coefficient                         & $3.3685 \cdot 10^{\!-\!6}$   &  L/s \T \\
$\epsilon$  			    & Porosity of carbon electrode                  & $0.87$		            &  -- \T \\
$E^0_{\text{cell}}$  	    & Equilibrium cell potential		                & $2.2$                     & V \T \B  \\
\hline
\hline
\end{tabular}
\vskip+0.1cm
\caption{{\small Parameters values for RFB experiment}}
\vskip-0.1cm
\label{tconst}
\end{table}
%-----------------------

\section{Concluding Remarks}
In this article, for an isothermal lumped parameter model of disproportionation redox flow batteries, an observer design for the simultaneous estimation of the battery states and crossover flux have been presented. The design considers an augmented space representation of the battery model and is based on Lyapunov stability theory. The observer linear feedback gain is obtained by solving a polytopic LMI problem, providing a systematic methodology where the EUUB convergence of the augmented state estimation error is guaranteed. The crossover term has been modelled-approximated as the output of a linear differential equation, enabling continual parametric estimation of the crossover flux as the battery discharges. Data from a vanadium acetylacetonate DRFB undergoing self-discharge was analyzed, demonstrating the performance of the observer proposed, in particular with respect to transient dynamic and noise rejection in the crossover flux estimation.

\section{Acknowledgments}
This work was carried out with funding support received from the Faraday Institution ({\tt faraday.ac.uk}; EP/S003053/1), grant number FIRG003.

%%%%%%%%%%%%%%%%%%%%%%%%%%%%%%%%%%%%%%%%%%%%%%%%%%%%%%%%%%%%%%%%%%%%%%%%%%%%%%%%
%----------------------------------------------------------------------------------------------------
\bibliographystyle{unsrt}
\bibliography{ASMH_CCTA2019_bib}
%----------------------------------------------------------------------------------------------------

\end{document}